\documentclass{IEEEtran}
\usepackage{cite}
\usepackage{amsmath,amssymb,amsfonts}
\usepackage{algorithmic}
\usepackage{graphicx}
\usepackage{color}
\usepackage{pdfpages}
\usepackage{bm}
\usepackage{array}
\usepackage{url}

\ifCLASSOPTIONcompsoc
    \usepackage[caption=false,font=footnotesize,labelfont=rm,textfont=rm]{subfig}
\else
\usepackage[caption=false, font=footnotesize]{subfig}
\fi
\usepackage{textcomp}

\def\BibTeX{{\rm B\kern-.05em{\sc i\kern-.025em b}\kern-.08em
    T\kern-.1667em\lower.7ex\hbox{E}\kern-.125emX}}

\newcommand{\iz}{{\bm i}_z}
\newcommand{\irho}{{\bm i}_\rho}
\newcommand{\iphi}{{\bm i}_\phi}

\newcommand{\normal}{{\bm i}_n}

\begin{document}
\title{Efficient 2.5-D FEM-Based Scattering Analysis of the Human Body for RF Sensing}
\author{Haoqing Wen, Michele D'Amico, \IEEEmembership{Senior Member, IEEE}, Matteo Oldoni, \IEEEmembership{Member, IEEE}, Federica Fieramosca,  \IEEEmembership{Graduate Student Member, IEEE}, Vittorio Rampa, \IEEEmembership{Senior Member, IEEE}, Stefano Savazzi, \IEEEmembership{Senior Member, IEEE}, Qi Wu, \IEEEmembership{Member, IEEE} and Gian Guido Gentili, \IEEEmembership{Member, IEEE}
\thanks{This work was supported in part by the China Scholarship Council program (Project ID: 202506020097), the National Science Foundation of China under grant 62525104, and was supported in part by the European Union under the Italian National Recovery and Resilience Plan (PNRR) of NextGeneration EU, partnership on “Telecommunications of the Future” (PE00000001 - program “RESTART”, Structural Project SRE). (\textit{Corresponding
author: Gian Guido Gentili.})}
\thanks{Haoqing Wen and Qi Wu are with the School of Electronics and Information Engineering,
Beihang University, Beijing 100083, China, and also with the Zhongguancun
Laboratory, Beijing 100094, China (e-mail: haoqing.wen@mail.polimi.it; qwu@buaa.edu.cn).}
\thanks{Gian Guido Gentili, Michele D’Amico, Matteo Oldoni, and H. Wen are with the Dipartimento di Elettronica, Informazione e Bioingegneria (DEIB), Politecnico di Milano, 20133 Milan, Italy (e-mail: gianguido.gentili@polimi.it; michele.damico@polimi.it; matteo.oldoni@polimi.it).}
\thanks{Stefano Savazzi, Vittorio Rampa, and Federica Fieramosca are with the Institute of Electronics, Computer and Telecommunication Engineering (IEIIT), National Research Council of Italy (CNR), 20133 Milan, Italy (e-mail: stefano.savazzi@cnr.it; vittorio.rampa@cnr.it; federica.fieramosca@cnr.it).}}

\maketitle

\begin{abstract}
Model training for Device-Free Localization (DFL) and Radio-Frequency (RF) sensing heavily relies on large-scale datasets, which are difficult, expensive, and time-consuming to obtain through measurements. This paper proposes a fast 2.5-dimensional Finite Element Method (2.5-D FEM) for computing the scattering fields of a Body of Revolution (BoR) human model under the excitation of a z-directed dipole. The proposed method can evaluate the effect of human micro-movements through the statistical characteristics of the Received Signal Strength Indicator (RSSI). The numerical accuracy and the practical applicability of the proposed method are validated through comparisons with full-wave simulations and indoor RF sensing experiments. The simulation results show agreement with the experimental measurements, demonstrating that the method is a reliable tool for evaluating micro-movement-induced statistical variations. The proposed method provides a practical and efficient means for generating large-scale, labeled RF training datasets, thereby accelerating the development of indoor localization tools as well as the calibration and tuning of tomographic reconstruction methods.
\end{abstract}

\begin{IEEEkeywords}
2.5-D finite element method, Body of Revolution, device-free localization, human scattering, RF sensing.
\end{IEEEkeywords}

\section{Introduction}

\IEEEPARstart{D}{evice}-Free Localization (DFL) or Radio-Frequency (RF) sensing have attracted significant attention in recent years due to their potential in a wide range of indoor applications, including smart environments~\cite{DFL_smartcity}, multi-target recognition~\cite{Vittorio202201,Vittorio202202}, health care~\cite{DFL_healthcare}, and human activity recognition~\cite{DFL_HumanActive} just to cite a few. Unlike conventional wireless systems that rely on active devices carried by users, DFL techniques, also known as passive methods, achieve target sensing and localization by exploiting the perturbations induced by human presence or motion on the surrounding electromagnetic field distributions and scattering characteristics~\cite{Fieramosca_awpl}, or on the wireless propagation channel~\cite{Stefano2019}. In indoor environments, the human body acts as a complex scatterer whose electromagnetic properties and spatial variations lead to measurable changes in the Received Signal Strength Indicator (RSSI), Channel State Information (CSI), or other RF observables \cite{review2017}. Accurate characterization of human-induced scattering effects is therefore a fundamental problem in the design and analysis of RF sensing systems.

\begin{figure}[!t]
\centerline{\includegraphics[width=2.7in]{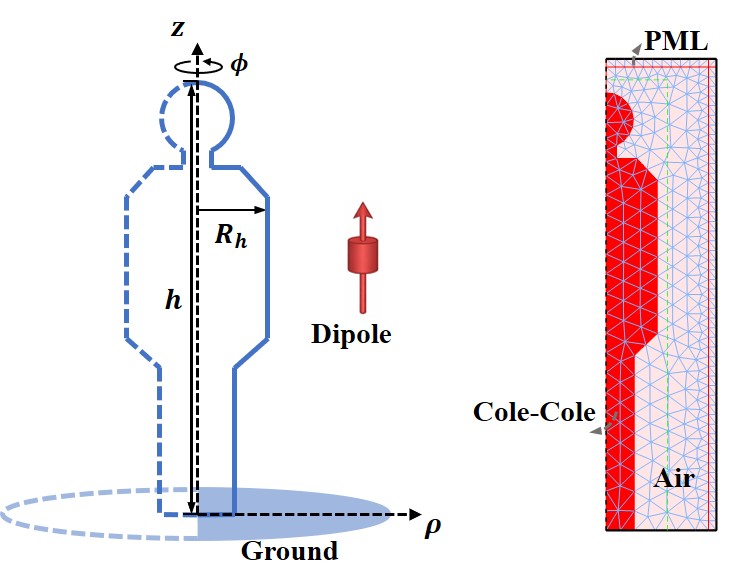}}
\caption{Schematic of the human body under the excitation of a $z$-directed  dipole, including an illustrative 2.5-D FEM mesh with labeled material regions (the mesh is shown for illustration only, a finer mesh is employed in the computations).}
\label{fig.body}
\end{figure}

With the widespread adoption of data-driven and machine learning techniques in RF localization and sensing, a variety of methods have been proposed, such as Bayesian inference~\cite{ML_Bayesian}, variational autoencoders (VAE)~\cite{ML_VAE_Stefano,ML_VAE_Federica}, generative neural network (GNN)~\cite{ML_GNN_Stefano,ML_GNN_Barba,ML_review}, to track human positions, movements, and activities. Most of these methods rely on approximate knowledge of a physical (prior) model to interpret the perturbations in radio signals induced by human presence or motion. Although tremendous efforts have been made to improve their accuracy and robustness, the performance of these models still relies heavily on training data covering diverse scenarios, body poses, positions, and states. However, collecting sufficiently large datasets typically requires extensive measurements across multiple individuals, different RF device deployments and setups, and several environmental conditions, which is costly, labor-intensive, and time-consuming. Therefore, efficient generation of high-quality RF databases for training and validation is still an open issue in machine-learning-based DFL and RF sensing.

To mitigate the difficulty of extensive measurement campaigns, EM modeling and numerical simulation techniques are often employed to generate prior datasets for network training. Existing approaches can be broadly categorized into three categories. Full-wave electromagnetic methods, such as the three-dimensional (3-D) Finite Element Method (FEM)~\cite{3DFEM}, Finite-Difference Time-Domain (FDTD)~\cite{FDTD01,FDTD02}, and Method of Moments (MoM)~\cite{MOM01,MOM02}, provide high modeling accuracy but require extraordinary computational time and resources, which limits their applicability to large-scale database generation. Analytical or semi-analytical models approximate the human body using canonical geometries, such as spheres \cite{GreenHead} or (infinitely long) cylinders \cite{AnaCylind01,AnaCylind02,AnaCylind03}, offering high efficiency at the cost of severely limited realism in representing human scattering characteristics. High-frequency approximation methods \cite{hf_method01,hf_method02,hf_method03,hf_method04} offer computational efficiency for large-scale propagation analysis, but lack accuracy in modeling near-field scattering and asymmetric fields from off-axis excitations. Consequently, existing simulation methods still face a trade-off between efficiency and accuracy, making them inadequate for generating large-scale, high-fidelity datasets required by machine-learning-based DFL systems.

Modeling the human body according to the BoR approach preserves its key features and makes the 2.5-D finite element method (2.5-D FEM) a suitable approach for human electromagnetic scattering. In fact, compared to conventional full 3-D solvers, 2.5-D FEM preserves the essential scattering characteristics of the body while, by exploiting the rotational symmetry of the structure, it significantly reduces the computational and memory footprint, enabling more efficient solutions of electromagnetic fields over large-scale parameter spaces~\cite{Jinjianming_borfem01,Jinjianming_borfem02,BOR_FEM,2.5DFEM02,2.5DFEM03,2.5DFEM04}. While 2.5-D FEM has been widely applied to a variety of electromagnetic problems, most existing formulations rely on symmetric excitations or uniform structures, which lack systematic analyses of off-axis excitations and realistic human models with complex scattering characteristics. Therefore, effectively extending 2.5-D FEM to simulate complex human scattering under off-axis excitations and validating its performance in DFL applications remains an important, but overlooked, challenge.

This paper proposes a fast 2.5-D FEM for computing the scattering fields of a BoR human body model under the excitation of a $z$-directed  dipole. The proposed approach enables efficient and accurate simulation of RF field distributions in indoor DFL and RF sensing scenarios. Using this method, large-scale synthetic datasets can be generated to train RF sensing models based on physical priors.

Specifically, the contributions of this work are threefold: 
\begin{itemize}
    \item \textit{Efficient 2.5-D FEM for dipole-excited BoR model:} a 2.5-D FEM formulation is developed for dipole excitation in BoR human models. Compared with conventional 3-D FEM, the proposed approach significantly reduces computational complexity while maintaining high numerical accuracy, making it well-suited for large-scale indoor electromagnetic simulations. 
    \item \textit{Validation for indoor RSSI prediction:} the effectiveness of the proposed method in predicting spatial RSSI distributions is validated through comparisons between simulated and measured results in an indoor DFL scenario. The results demonstrate good agreement, confirming the reliability of the method for practical RSSI-based applications.
    \item \textit{Synthetic RF dataset generation tool:} by modeling human micro-movements and analyzing the resulting RSSI statistics, the method is shown to generate synthetic data with statistical properties consistent with measurements, enabling efficient production of labeled RF training datasets. 
\end{itemize}


The rest of the paper is organized as follows. Sect.~\ref{chap:method} presents the theoretical formulations of the 2.5-D FEM for the electromagnetic scattering computation of a human body under an  $z$-directed dipole excitation. Sect.~\ref{chap:numerical_results} validates the proposed method and provides numerical results in a DFL scenario. In Sect.~\ref{chap:experimental_results}, the simulation results are compared with measurements to evaluate the accuracy and the applicability of the method. Finally, Sect.~\ref{chap:conclusion} draws some conclusions.

\section{Numerical Method}
\label{chap:method}

The use of a BoR to represent the human body provides an excellent trade-off between accuracy and computational speed. The evaluation of typical parameters in DFL scenarios requires a large number of simulations, thus efficiency plays a key role in the overall applicability of the method. The BoR approximation allows to decouple the contribution of different harmonics in which the field can be decomposed, thereby significantly reducing the computational time spent solving the associated linear system of equations derived from the FEM application. Perfectly matched Layer (PML) is used in proximity of the BoR to accurately simulate free space, so that the domain of FEM computation is limited to a rather small region around the BoR. In order to compute the field at any distance from the BoR, we deploy equivalent sources all around the body and use them for field computation outside of the FEM domain. This also contributes to the significant speed of analysis of the overall methodology.

\subsection{2.5-D FEM}

In this section, we start discussing the FEM discretization and proceed with the post-processing derived from the equivalent sources. The problem we want to solve here is that of an arbitrarily shaped BoR placed over an infinite half-space and illuminated by a vertically directed Hertzian dipole. The BoR is centered in a local cylindrical reference frame  $(\rho,\phi,z)$, with unit vectors $\irho$, $\iphi$, $\iz$. The source that provides the forcing term is a $z$-directed Hertzian dipole placed outside the FEM domain. With no loss of generality, the source position is set at ${\bf r}_s=(\rho_s,\phi_s, z_s)$.

Within the scattered field formulation of the 2.5-D FEM problem, the scattered electromagnetic field is expanded as a sum of circular harmonics in $\phi$ and solved in the $\rho, z$ plane. In that plane, we label ${\bf E}_t$ the in-plane component of the electric field ($\rho$-$z$ components) and ${\bf E}_\phi$ the out-of-plane component. Using the superscript $(m)$ to refer to the generic harmonic in $\phi$, with $M$ the number of harmonics, we can write the following expansion for the scattered electric field ${\bf E}$ as
\begin{equation}
    {\bf E}(\rho,\phi,z) = {\bf E}_t(\rho,\phi,z)+ {\bf E}_\phi(\rho,\phi,z),
    \label{eq:Et_Ephi}
\end{equation}
with
\begin{align}
    {\bf E}_t(\rho,\phi,z) = \sum_{m=0}^M{\bf e}_t^{(m)}(\rho,z)c_m, \\
 {\bf E_\phi}(\rho,\phi,z) = \iphi\sum_{m=1}^M {e}_\phi^{(m)}(\rho,z)s_m,
\end{align}
where 
\begin{equation}
c_m = \frac{1}{\sqrt{\pi\varepsilon_{m}}}\cos m(\phi-\phi_s),
\end{equation}
\begin{equation}
s_m = \frac{1}{\sqrt{\pi}}\sin m(\phi-\phi_s),
\end{equation}
and 
\begin{equation}
\varepsilon_{m}=\left\{
\begin{array}{ll}
2 & \text{if } m=0 \\
1 & \text{if } m> 0
\end{array}.
\right.
\end{equation}
The specific $\phi$ dependency of the in-plane and out-of-plane components is dictated by the direction of the source and by its position.

We can now write the weak form of the scattered field formulation of FEM \cite{Jinbook} in the domain $V$ for penetrable scatterers, in which the forcing term is the field of the dipole (outside of the FEM domain). This is expressed as
\begin{align}\label{eq:weak_form}
        \int_V \nabla\times{\bf W}\cdot\nabla\times{\bf E}\; dV - k_0^2\int_V \epsilon_r {\bf W}\cdot{\bf E}\;dV +\nonumber \\
        j\omega\mu_0\int_{\partial V_\text{gnd}} {\bf W}\cdot({\bf H}\times \normal)       dS =k_0^2\int_{V} (\epsilon_r - 1){\bf W}\cdot{\bf E}_i \;dV,
\end{align}
where ${\bf E}$ is the scattered electric field, ${\bf H}$ is the corresponding magnetic field, ${\bf W}$ is an arbitrary testing function belonging to $H(\text{curl})$ and satisfying essential boundary conditions, ${\bf E}_i$ is the incident field, $\epsilon_r$ is the complex relative permittivity of the human body, $\partial V_\text{gnd}$ is the part of the boundary corresponding to the ground with $\partial V_\text{gnd}\subset \partial V$, $k_0$ is the free-space wavenumber and $\normal$ is the outward directed normal unit vector. Using the local cylindrical reference frame, we find
\begin{equation}
    \int_V (.) \, dV = \int_0^{2\pi}\!\!d\phi\,\!\!\int_S(.)\,\rho \, d\rho \, dz   ,
\end{equation}
where $S$ is a cut plane described by the coordinates $\rho,z$ and corresponding to the $\phi=\phi_s$ plane. Thanks to the BoR approximation for the scatterer, the integral in $\phi$ is obtained analytically, as seen in the following, and the FEM domain is then limited to $S$ only.

To discretize (\ref{eq:weak_form}), we used Galerkin's method and a mixed finite-element basis. The in-plane component ${\bf e}_t^{(m)}$ is expanded using edge-based basis functions $\left\{{\bm\tau}_q(\rho,z)\right\}$, whereas the function ${\rho e_\phi^{(m)}}$ is expanded using Lagrange nodal basis functions $\left\{\varphi_{q}(\rho,z)\right\}$ \cite{Jinjianming_borfem01,Jinjianming_borfem02}. So the two components of (\ref{eq:Et_Ephi}) can be expressed as
\begin{equation}\label{eq:Et}
    {\bf e}_t^{(m)}(\rho,z) = \sum_{q=1}^{Q} u_{q}^{(m)}{\bm\tau}_q(\rho,z), 
\end{equation}
\begin{equation}\label{eq:Ephi}
    \rho e_\phi^{(m)}(\rho,z) =  \sum_{q=1}^{Q'}v_{q}^{(m)}\varphi_{q}(\rho,z),
\end{equation}
where $Q$ and $Q'$ denote the numbers of basis functions associated with ${\bf e}_t^{(m)}$ and $\rho e_\phi^{(m)}$, respectively. 

Using the subscript '$t$' for the in-plane component and '$\phi$' for the out-of-plane component, from the expansions (\ref{eq:Et})-(\ref{eq:Ephi}), we can verify that
\begin{align}
    & (\nabla\times{\bm\tau}c_m)_t = -\frac{m}{\rho}\left(\tau_z\irho - \tau_\rho \iz\right) s_m,
    \label{eq:curl_vec_t}\\
    & (\nabla\times{\bm\tau}c_m)_\phi = \left(\frac{\partial\tau_\rho}{\partial z}-\frac{\partial \tau_z}{\partial \rho}\right)c_m ,
    \label{eq:curl_vec_p}    
\end{align}
\begin{equation}\label{eq:curl_sca}
    \nabla \times \left(\iphi\frac{\varphi}{\rho }s_m\right)= 
\frac{(\nabla \times \iphi\varphi)_t}{\rho} s_m,
\end{equation}
\begin{equation}\label{eq:nablavarphi}
    (\nabla\times\iphi\varphi)_t = 
    \frac{\partial\varphi}{\partial\rho}\iz - \frac{\partial\varphi}{\partial z}\irho.
\end{equation}

In Galerkin's approach, the set of testing functions is the same set of basis functions, so, in addition to the full set of expansion functions
\begin{align}
  {\bf E}_t\rightarrow  \left\{{\bm\tau}_q c_m\right\},\hspace{1em}m=0\ldots M,\hspace{1em} q=1\ldots Q \\
\rho{E}_\phi\rightarrow    \left\{{\varphi}_q s_m\right\},\hspace{1em}m=1\ldots M,\hspace{1em}q=1\ldots Q',
\end{align}
we introduce the sets of testing functions
\begin{align}
  {\bf W}_t\rightarrow  \left\{{\bm\tau}_p c_n\right\},\hspace{1em}n=0\ldots M,\hspace{1em} p=1\ldots Q \\
\rho{W}_\phi\rightarrow    \left\{{\varphi}_q s_n\right\},\hspace{1em}n=1\ldots M,\hspace{1em}p=1\ldots Q',
\end{align}
and we test the weak form using the aforementioned set. We obtain the following sparse system of equations
\begin{equation}\label{eq:main}
    \left(\mathbb{A}-k_0^2\mathbb{B} + j\omega\mu_0\mathbb{C}\right)\mathbb{U} = k_0^2\mathbb{K},
\end{equation}
where $\mathbb{A}$ is block-diagonal 
\begin{equation}
\mathbb{A} =
\begin{bmatrix}
\mathbb{A}^{(0)} & 0 & \cdots & 0 \\
0 & \mathbb{A}^{(1)} & \cdots & 0 \\
\vdots & \vdots & \ddots & \vdots \\
0 & 0 & \cdots & \mathbb{A}^{(M)}
\end{bmatrix},
\end{equation}
$\mathbb{B}=\mathbb{I}\otimes\mathbb{B}'$ and $\mathbb{C}=\mathbb{I}\otimes\mathbb{C}'$ with $\otimes$ being the Kronecker product, $\mathbb{I}$ the identity matrix of size $M+1$ and
\begin{equation}
    \mathbb{U} =     
    \begin{bmatrix}
    \mathbb{U}^{(0)} \\
    \mathbb{U}^{(1)} \\
    \vdots   \\
    \mathbb{U}^{(M)}\\
    \end{bmatrix}, \hspace{2em}
    \mathbb{K} =     
    \begin{bmatrix}
    \mathbb{K}^{(0)} \\
    \mathbb{K}^{(1)} \\
    \vdots   \\
    \mathbb{K}^{(M)}\\
    \end{bmatrix}.
\end{equation}

The block-diagonal structure of the system in \eqref{eq:main} is owing to the fact that
\begin{equation}
    \int_0^{2\pi} c_m c_n d\phi = 
    \left\{
    \begin{array}{ll}
    1 &\text{if }m=n \\
    0 &\text{if }m\neq n \\    
    \end{array}
    \right. ,
\end{equation}
\begin{equation}
    \int_0^{2\pi} s_m s_n d\phi = 
    \left\{
    \begin{array}{ll}
    1 &\text{if }m=n>0 \\
    0 &\text{if }m\neq n \\    
    \end{array}
    \right. ,
\end{equation}
\begin{equation}
    \int_0^{2\pi} c_m s_n d\phi = 0.
\end{equation}
We can further express the individual matrices of the system as
\begin{equation}\label{eq:Amat}
    \mathbb{A}^{(m)} = \begin{bmatrix}
        \mathbb{A}_{tt} + m^2\mathbb{A}'_{tt} & -m\mathbb{A}_{t\phi} \\
        -m\mathbb{A}_{\phi t} & \mathbb{A}_{\phi\phi} \\       
    \end{bmatrix} ,
\end{equation} 
\begin{equation}
\mathbb{B}' =\begin{bmatrix}
         \mathbb{B}_{tt} & {0} \\
        {0} & \mathbb{B}_{\phi\phi} \\       
\end{bmatrix} ,
\end{equation}
\begin{equation}
\mathbb{C}' =
\begin{bmatrix}\label{eq:knownterm}
        \mathbb{C}_{tt} & 0\\ 0 & \mathbb{C}_{\phi\phi} 
    \end{bmatrix},
\end{equation}    
\begin{equation}
        \mathbb{U}^{(m)} = \begin{bmatrix}\mathbb{U}_{t}^{(m)} \\ \mathbb{U}_{\phi}^{(m)} 
    \end{bmatrix},\hspace{1em}
\mathbb{K}^{(m)} = \begin{bmatrix}
        \mathbb{K}_{t}^{(m)} \\ \mathbb{K}_{\phi}^{(m)} 
    \end{bmatrix},
\end{equation}
and 
\begin{equation}
    \mathbb{U}_t^{(m)} = [u_1^{(m)}\; u_2^{(m)} \ldots u_{Q}^{(m)}]^\text{T},
\end{equation} 
\begin{equation}
\mathbb{U}_\phi^{(m)} = [v_1^{(m)}\; v_2^{(m)} \ldots v_{Q'}^{(m)}]^\text{T},
\end{equation} 
where superscript T indicates transpose.
The explicit expressions of the individual terms of matrices $\mathbb{A }^{(m)}$, $\mathbb{B}'$, $\mathbb{C}'$ are provided in the Appendix \ref{app:matrix_element}. In the following section, we will show in more detail the computation of the known term.

From system (\ref{eq:main}), thanks to the block diagonal structure of the matrices involved, we obtain
\begin{equation}\label{eq:main_m}
    \mathbb{U}^{(m)} = k_0^2\left(\mathbb{A}^{(m)}-k_0^2\mathbb{B}' + j\omega\mu_0\mathbb{C}'\right)^{-1}\mathbb{K}^{(m)}.
\end{equation}
Because the matrices are relatively small, a direct solver is used to obtain the solution in this work. The advantage of using the BoR representation is twofold:
\begin{itemize}
    \item the block diagonal structure of the matrices involved, allowing a huge reduction of computation time. A detailed study of the computational complexity is well beyond this work, but we estimate that the solution time for a full 3D FEM discretization is at least $M^2$ times larger. A typical value of $M$ is 10, so a gain factor of at least 100 is expected.
    \item The use of a 2D mesh. This is another advantage, as the time required to create the mesh is negligible. 
\end{itemize}

\subsection{Known term: $z$-directed Hertzian dipole}

The method can handle a quite arbitrary source, but, in order to match the practical transmitting and receiving antennas in DFL scenarios, an arbitrarily placed $z$-directed hertzian dipole was chosen as the excitation source. Such a source provides the incident electric field ${\bf E}_i$ of the scattered field formulation of FEM.  Without loss of generality, the hertzian dipole is assumed to be located at ${\bf r}_s =(\rho_s,0,z_s)$, with a source moment ${\bf p} = \iz Il$. From the knowledge of the power $P_r$ radiated by the source, we can obtain the source moment component (assumed real) as
\begin{equation}
    Il = \sqrt{\frac{3P_r\lambda_0^2}{\pi\eta_0}},
\end{equation}
where $\lambda_0$ is the  free-space wavelength and $\eta_0 = 376.73$ $\Omega$ is the free-space impedance. The presence of a partially reflecting ground (i.e., geometrically represented by the plane $z=0$) with unknown material parameters prevents the use of an exact expression for the field radiated by the dipole. The contribution of the reflecting ground has been approximated by an image source with a moment $\tilde{\bf p} = \Gamma{\bf p}$, placed at $\tilde{\bf r}_s = (\rho_s,\phi_s,-z_s)$. Assuming that the dipole is not very close to the human body (i.e, $R\gg \lambda_0$), we can write the total incident field at the generic position ${\bf r}$ as: 
\begin{equation}\label{eq:zdipole_close}
\mathbf{E}_i(\mathbf{r})
= -j\omega\mu_0 \Bigg[
 \frac{e^{-jkR}}{4\pi R}\,
\mathbf{p}_\perp 
 + \frac{e^{-jkR'}}{4\pi \tilde{R}}\,
\tilde{\mathbf{p}}_\perp  
\Bigg],
\end{equation}
where $R = |{\bf r}-{\bf r}_s|$. and $\tilde{R} = |{\bf r}-\tilde{\bf r}_s|$. The subscript '$\perp$' is a compact form for
\begin{equation}\label{eq:perp1}
    {\bf p}_\perp = \frac{{\bf r}-{\bf r}_s}{R}\times{\bf p}\times\frac{{\bf r}-{\bf r}_s}{R},
\end{equation}
\begin{equation}\label{eq:perp2}
    \tilde{\bf p}_\perp = \frac{{\bf r}-\tilde{\bf r}_s}{\tilde{R}}\times\tilde{\bf p}\times\frac{{\bf r}-\tilde{\bf r}_s}{\tilde{R}}.
\end{equation}

The field expressions thus obtained are now decomposed into azimuthal Fourier modes in $\phi$. We can write 
\begin{equation}
    {\bf E}_i = \sum_{m=0}^M {\bf e}_{i,t}^{(m)}c_m + \iphi\!\sum_{m=1}^M {e}_{i,\phi}^{(m)}s_m,
\end{equation}
where ${\bf e}_{i,t}^{(m)} = \irho e_{i,\rho}^{(m)} + \iz e_{i,z}^{(m)}$, to find
\begin{align}
    & e_{i,\rho}^{(m)}(\rho,z)=  \int_0^{2\pi} \irho\cdot{\bf E}_i({\bf r}) c_m d\phi,\label{eq:Er_m}\\
    & e_{i,z}^{(m)}(\rho,z)=  \int_0^{2\pi} \iz\cdot{\bf E}_i({\bf r}) c_m d\phi,\label{eq:Ez_m}\\
    & e_{i,\phi}^{(m)}(\rho,z)=\int_0^{2\pi}\iphi\cdot{\bf E}_i({\bf r})s_m d\phi.\label{eq:Ep_m}
\end{align}
From the decomposition in Fourier harmonics, we can obtain the expression of the known term in (\ref{eq:knownterm}). We have
\begin{equation}
    \mathbb{K}_t^{(m)}(p) = \int_S (\epsilon_r - 1) {\bm\tau}_p\cdot {\bf e}_{i,t}^{(m)}\rho d\rho dz,
\end{equation}
\begin{equation}
    \mathbb{K}_\phi^{(m)}(p) = \int_S (\epsilon_r - 1) {\varphi}_p { e}_{i,\phi}^{(m)}d\rho dz.
\end{equation}

\subsection{External field computation by equivalent sources}

The solution of (\ref{eq:main_m}) yields the scattered electric and magnetic  fields everywhere in the FEM domain as
\begin{equation}\label{eq:fieldFEME}
    {\bf E} = \sum_{m=0}^M\sum_{q=1}^Q u_{q}^{(m)}{\bm\tau}_qc_m + \iphi\sum_{m=1}^M\sum_{q=1}^{Q'} v_{q}^{(m)}\frac{\varphi_q}{\rho} s_m,
\end{equation}
\begin{align}\label{eq:fieldFEMH}
    {\bf H} =& \left\{\sum_{m=0}^M\sum_{q=1}^Q u_{q}^{(m)}\left[\iphi(\nabla\times{\bm\tau}_q)_\phi c_m-\frac{m}{\rho} {(\nabla\times{\bm\tau}_q)_t}s_m\right]  \nonumber \right. \\ 
    & +\left. \sum_{m=1}^M\sum_{q=1}^{Q'} v_{q}^{(m)}{(\nabla\times\iphi\varphi_q)_t} s_m \right\}\frac{1}{-j\omega\mu_0},
\end{align}
where, to compact the notation, we set
\begin{equation}
    (\nabla\times\bm\tau)_t = {\tau_x\irho -\tau_\rho\iz},
\end{equation}
\begin{equation}
    (\nabla\times\bm\tau)_\phi =\frac{\partial\tau_\rho}{\partial z}-\frac{\partial \tau_z}{\partial \rho} 
\end{equation}
and where $(\nabla\times\iphi\varphi)_t$ is given by (\ref{eq:nablavarphi}).
\begin{figure}[!t]
\centerline{\includegraphics[width=2.5in]{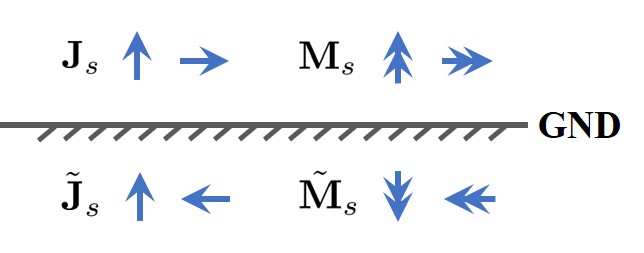}}
\caption{A sketch of the images used for the equivalent electric and magnetic currents.}
\label{fig.image}
\end{figure}

Starting from (\ref{eq:fieldFEME})--(\ref{eq:fieldFEMH}), the scattered fields at all points outside the FEM domain can be computed by using equivalent sources. This post-processing solves the issue of discretizing by FEM a large space around the human body to compute the field at all receivers in the environment, which can be placed at a significant distance from the human body. A cylindrical surface $S_\text{eq}$ in proximity of the human body is defined, and the scattered fields on this surface are computed by (\ref{eq:fieldFEME})--(\ref{eq:fieldFEMH}). An outward normal unit vector $\normal$ is introduced on $S_\text{eq}$ and surface equivalent electric (${\bf J}_s$) and magnetic (${\bf M}_s$) current distributions are deployed on $S_\text{eq}$. They are related to the fields by 
\begin{equation}\label{eq:JM}
    {\bf J}_s = \normal\times{\bf H},
\end{equation}
\begin{equation}\label{eq:M}
     {\bf M}_s = -\normal\times{\bf E}.
\end{equation}
The electric field at all points outside $S_\text{eq}$ is then computed by
\begin{align}\label{eq:eqsources}
    {\bf E}({\bf r}) = -j\omega\mu\int_{S_\text{eq}}\left({\bf J}_{s\perp}\frac{e^{-jkR}}{4\pi R}dS' + 
 \tilde{\bf J}_{s\perp}\frac{e^{-jk\tilde{R}}}{4\pi \tilde{R}}\right)dS'  \nonumber \\  + jk\int_{S_\text{eq}}\left(\frac{\bf R}{R}\times{\bf M}_{s}\frac{e^{-jkR}}{4\pi R} +
\frac{\tilde{\bf R}}{\tilde{R}}\times\tilde{\bf M}_{s}\frac{e^{-jk\tilde{R}}}{4\pi \tilde{R}} \right)dS',
\end{align}
where, subscript '$\perp$' was defined in (\ref{eq:perp1})--(\ref{eq:perp2}), $\tilde{\bf J}_s$ and $\tilde{\bf M}_s$ are image contributions, ${\bf R} = {\bf r}-{\bf r}'$, $R=|{\bf R}|$, $\tilde{\bf R}={\bf r}-\tilde{\bf r}'$, $\tilde{R} = |\tilde{\bf R}|$, being ${\bf r}'$ the source position and $\tilde{\bf r}'$ the image position. Similarly to what we did for the known term, we use a multiplicative coefficient $\Gamma$ for the image contribution and a direction that is synthetically represented in Fig. \ref{fig.image}. Expression (\ref{eq:eqsources}) is only a good approximation for two reasons
\begin{itemize}
    \item it only represents the field at a few wavelengths from the equivalent sources. This was considered a completely acceptable approximation because the receivers are generally at several wavelengths apart;
    \item the image contribution is approximated by an angle-independent coefficient $\Gamma$. 
\end{itemize}
The second hypothesis accounts for the large uncertainties in the composition of the ground in a building. The material can be far from homogeneous (e.g., rods and metal supports can be present), and the value of relative permittivity is unknown. However, in the measured data, an interference pattern from the ground was observed, and a synthetic average contribution of ground reflection was condensed in a single parameter $\Gamma$ \cite{3GPP_TR38901_v19_2025,Indoor_channel}. This can account for ground interference, with some agreement with measured values, and was considered acceptable in the modeling.

\begin{figure}[!t]
\centerline{\includegraphics[width=3in]{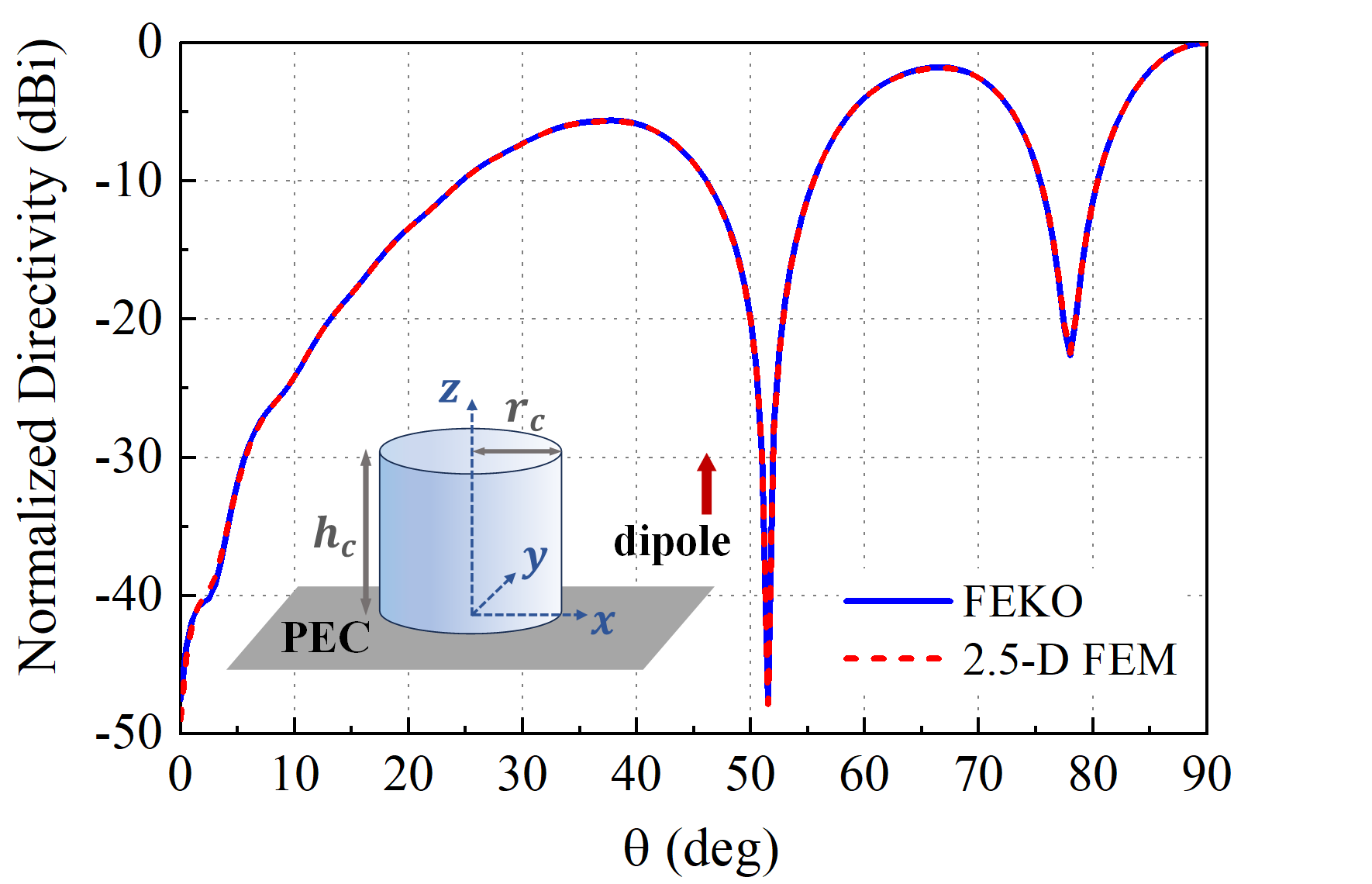}}
\caption{Normalized directivity at 2.43 GHz in $\phi=0$ plane for  $z$-directed electric dipole in the presence of a dielectric cylinder ($\epsilon_r = 52.7 -j12.76$) and a PEC ground plane.  The dipole is located at $\mathbf{r}_s = (1, 0, 0.15)$ m, the cylinder has height $h_c = 0.2$ m and radius $r_c = 0.1$ m. Comparison between 2.5-D FEM in this work and Feko$^\text{TM}$ simulations.}
\label{fig.vali_feko}
\end{figure}

\section{Validation and Results of 2.5-D FEM}
\label{chap:numerical_results}

We discuss in this section first a validation of the model and then its application to a real-case scenario, in which a person is present in a room where a set of transmitters and receivers are deployed all around. Our final purpose is to assess the usefulness of the model in a real environment, where several unknown effects are present. Thus, in the following sections, we define some useful statistical parameters that help the analysis.

\subsection{Model validation}

In the following discussion, we used the human body model shown in Fig.~\ref{fig.body}, where $h$ and $R_h$ are control parameters for the body height and girth of the human model, while the parameters of other body parts are scaled proportionally to these two values to ensure reasonable body proportions. In this work, the dielectric properties of human tissues have been obtained by the Cole–Cole model \cite{cole-cole}, in which the 6-term model with 14 parameters is used
\begin{equation}\label{eq:cole}   \epsilon_r(\omega)=\epsilon_{r,\infty}+\sum_{i=1}^4\frac{\Delta\epsilon_{r,i}}{1+\left({j}\omega\tau_i\right)^{1-\alpha_i}}-j\frac{\sigma}{\omega\epsilon_0}
\end{equation}
while the model parameters were obtained from \cite{cole-cole}, Table 1, assuming muscle as a homogeneous human tissue.
From the model, at the operating frequency of 2.43 GHz, the material has a complex relative permittivity $\epsilon_r = 52.7-{j}12.76$. 

    A complete validation of the human body model with Feko is cumbersome, as it is based on the MoM with different meshing and computation strategies, so we opted for a simpler model, consisting of a cylinder on a PEC ground plane. The cylinder has a base radius $r_c = 0.1$ m, a height $h_c = 0.2$ m, and is made of muscle tissue modeled using the Cole-Cole model previously introduced.  A $z$-directed hertzian dipole acting as a source term is located at $\mathbf{r} = (1, 0, 0.1)$ m.  This is the model used for validation as well, as it contains the most important features of the analysis we want to carry out. 

The results obtained are shown in Fig.~\ref{fig.vali_feko}, where the comparison of the normalized directivity computed by the proposed method and by Feko$^\text{TM}$ at $2.43~\mathrm{GHz}$ \cite{FEKO2025} is presented. The far field has been obtained by asymptotic expressions obtained from (\ref{eq:JM}) with $\Gamma = 1$ (exact image contribution). As shown in the figure, the two curves fit well, and we consider this analysis a sufficient validation of the accuracy of the proposed 2.5-D FEM fast computation method. 

The maximum harmonic index $M$ used in the analysis was 11. The FEM discretization corresponds to 2439 triangles and 17344 degrees of freedom (with 2nd order mixed elements \cite{2.5DFEM04}) for each harmonic in $\phi$. The procedure has been implemented in MATLAB, and the CPU time to solve the problem (relative to all harmonics) amounts to 4.6 sec. on a PC equipped with an Intel i9 processor working at 3.2 GHz and with 128 GB RAM, as compared to 7 sec. using Feko. For this case, the speed improvement is not really significant, but larger problems indicate a huge speed-up in the analysis, as shown in the following section.

\begin{figure}[!t]
\centerline{\includegraphics[width=3.2in]{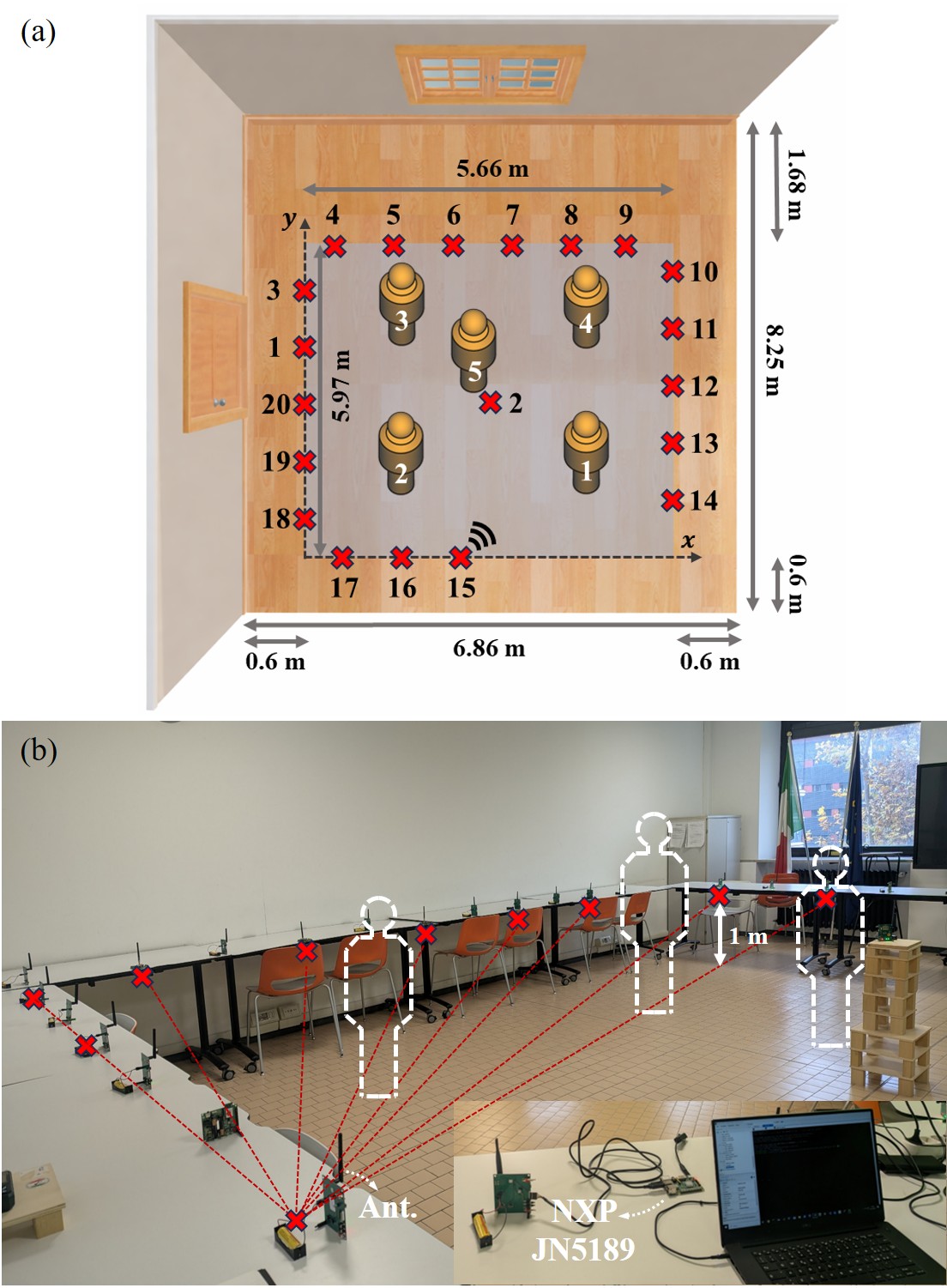}}
\caption{
(a) Top-view schematic diagram of the indoor DFL scenario. The gray rectangular region indicates the test area for antennas and human targets. Twenty antennas are deployed as transmitters/receivers, and five human positions are marked. The room entrance is in the lower right corner, where no antennas are deployed. (b) Illustration of the indoor DFL test scenario as seen from the room entrance. The Rx/Tx deployment is marked by red crosses. Human body positions are reported in dashed white lines. The RF acquisition setup is sketched on the lower right part of this figure.}
\label{fig.sys2}
\end{figure}

\begin{table}[t]
\caption{Relative coordinates $(x,y)$ of Antennas and Nominal Locations of the Human Body}
\label{tab:ant_human_coord}
\centering
\renewcommand{\arraystretch}{1.15}
\begin{tabular}{>{\centering\arraybackslash}p{0.65cm} c >{\centering\arraybackslash}p{0.65cm} c >{\centering\arraybackslash}p{0.65cm} c}
\hline\hline
Ant. & Coord. (m) & Ant. & Coord. (m) & Body & Coord. (m) \\
\hline
1  & (0.00, 3.51) & 11 & (5.66, 4.47) & 1 & (4.12, 1.97) \\
2  & (2.85, 3.43) & 12 & (5.66, 3.47) & 2 & (1.48, 1.97) \\
3  & (0.00, 4.51) & 13 & (5.66, 2.47) & 3 & (1.48, 4.61) \\
4  & (0.56, 5.97) & 14 & (5.66, 1.47) & 4 & (4.13, 4.61) \\
5  & (1.56, 5.97) & 15 & (2.50, 0.00) & 5 & (2.85, 3.43) \\
6  & (2.56, 5.97) & 16 & (1.50, 0.00) &   &              \\
7  & (3.56, 5.97) & 17 & (0.50, 0.00) &   &              \\
8  & (4.56, 5.97) & 18 & (0.00, 0.51) &   &              \\
9  & (5.56, 5.97) & 19 & (0.00, 1.51) &   &              \\
10 & (5.66, 5.47) & 20 & (0.00, 2.51) &   &              \\
\hline\hline
\end{tabular}
\end{table}

\begin{figure}[!t]
\centerline{\includegraphics[width=3.1in]{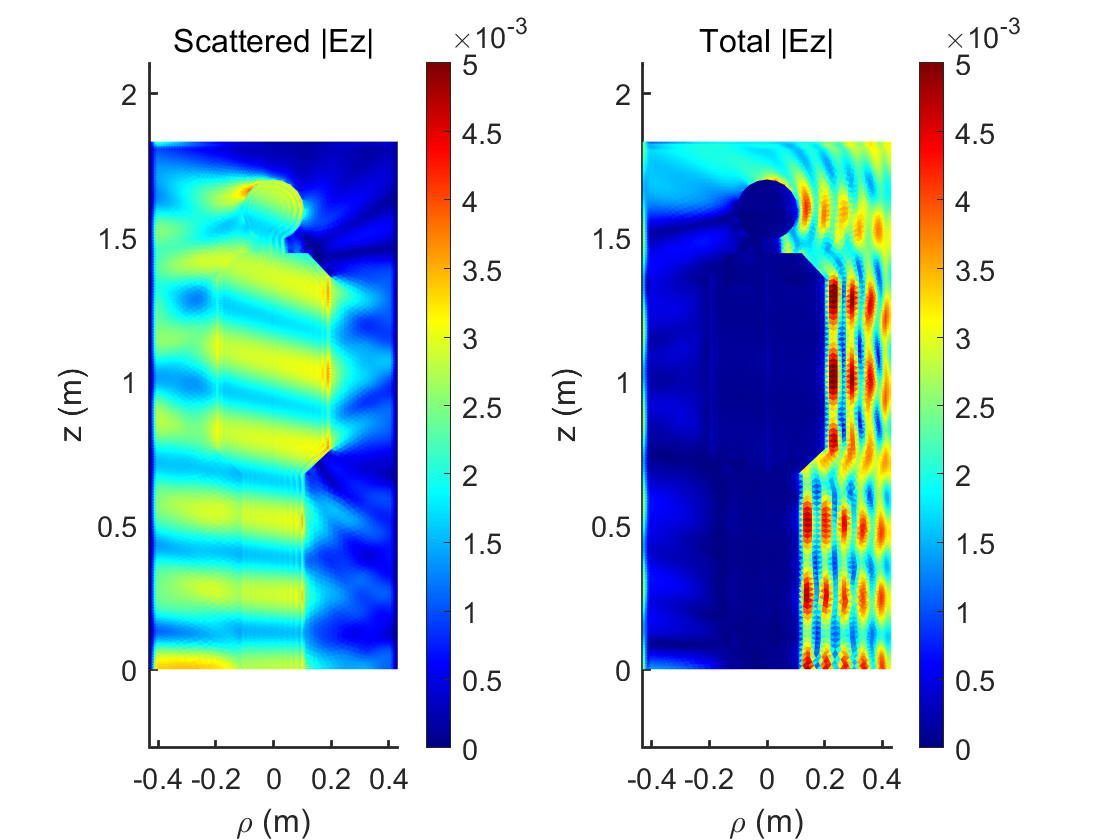}}
\caption{Scattered (left) and total (right) $E_z$ fields on the $\rho,z$ plane of the simulated human body, shown of Fig.~\ref{fig.body}, that is modeled by the Cole-Cole muscle model (\ref{eq:cole}) at 2.43 GHz, Tx = 7, and $p = 1$. The figure refers to the $\phi=\phi_s$ plane, and the source is located at the right in the figure.}
\label{fig.nearfield}
\end{figure}

\begin{figure*}[!t]
\centerline{\includegraphics[width=\textwidth]{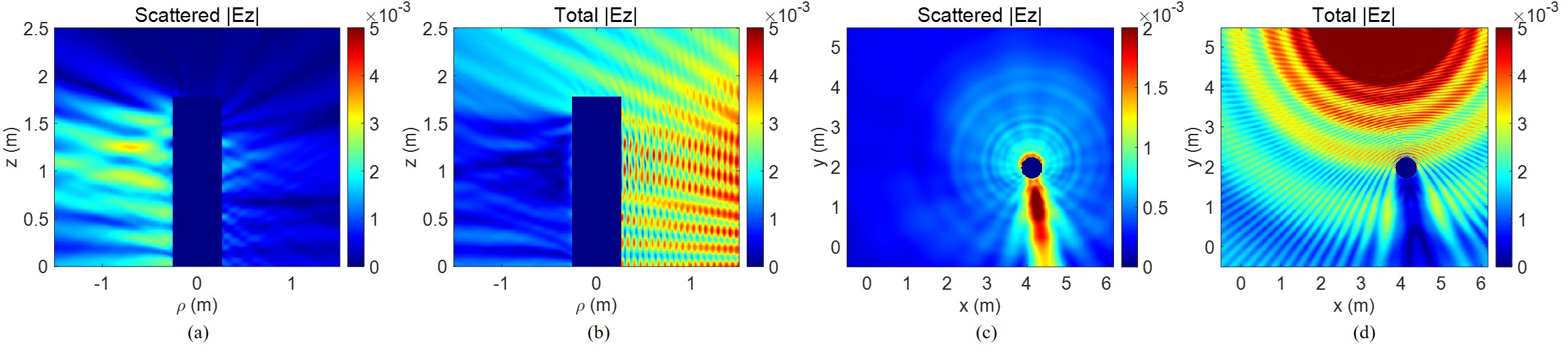}}
\caption{From left to right. $E_z$ fields computed by the equivalent sources for at 2.43 GHz for Tx = 7 and $p = 1$: (a) scattered field on the $\rho,z$ plane, (b) total field on the $\rho,z$ plane, both for $\phi = \phi_s$, (c) scattered field on the $x,y$ plane at $z=1$ m, and (d) total field on the $x,y$ plane at $z=1$ m, the Tx height over the ground. In all figures, the equivalent sources are slightly inside the dark region.}
\label{fig.Ez_rz_xy}
\end{figure*}


\section{Experimental setup and  comparison results}
\label{chap:experimental_results}

In this section, an indoor DFL scenario is introduced, and the field distributions and statistical characteristics are systematically analyzed. The scattered and total fields are computed, with the results compared with experimental measurements to validate the proposed method. Furthermore, human micro-movements are considered to investigate their influence on the statistical properties of the RSSI, providing a foundation for subsequent modeling and data generation based on these statistics.

\subsection{Scattered and Total Field Analysis}

An indoor DFL scenario was constructed to demonstrate the capabilities of the proposed approach to model a real-case study. The DFL environment is modeled as an indoor space with dimensions illustrated in Fig.~\ref{fig.sys2}(a). Twenty radio transceiver devices (with antennas labeled as $N = 1,\dots,20$) are arranged along a rectangular layout within the room, with the antennas uniformly distributed along each side of the rectangle and one antenna (i.e., antenna 2) near the center of the room. Five possible human body locations (labeled as $p = 1,\dots,5$) are considered in the scenario as detailed in the same figure. Taking the lower-left corner of the rectangle as the origin of the $x,y$ plane, the coordinates of the antennas and the predefined human positions are listed in Table~\ref{tab:ant_human_coord}. All antennas are arranged at 1 m above the ground. 

In our simulations, in order to better approximate a realistic indoor environment, the ground image coefficient is set to $\Gamma = 0.3$, while the effects of the walls, ceiling, and multipath propagation are neglected at the present stage. A typical human body model is used, with $h = 1.7$ m and $R_h = 0.2$ m. 
The FEM discretization corresponds to 30770 triangles and 216507 degrees of freedom (with 2nd order mixed elements \cite{2.5DFEM04}) for each harmonic in $\phi$, with a total number of harmonics $M=11$. The procedure has been implemented in MATLAB, and the CPU time to solve the problem (relative to all harmonics) amounts to 60 sec. on the same PC equipped with an Intel i9 processor previously cited. The post-processing that employs equivalent sources to compute the field at all receivers is almost instantaneous. The analysis with Feko was impractical because of the extremely large computing time (several hours).

As an example, the scattered and total $E_z$ field (the field component relevant at the receivers) is computed and plotted for the case of Tx = 7 and $p = 1$. As shown in Fig.~\ref{fig.nearfield}, the total electric fields clearly delineate the human body contour, indicating that the field penetrates very little inside the human body. A significant interference pattern is present in both the backward direction and the forward direction with respect to the incident field (arriving from the right in the figure). The total field also shows a pronounced shadow region behind the body and noticeable wavefront bending in front, with local field enhancement. 


\begin{figure*}[!t]
\centerline{\includegraphics[width=6.8in]{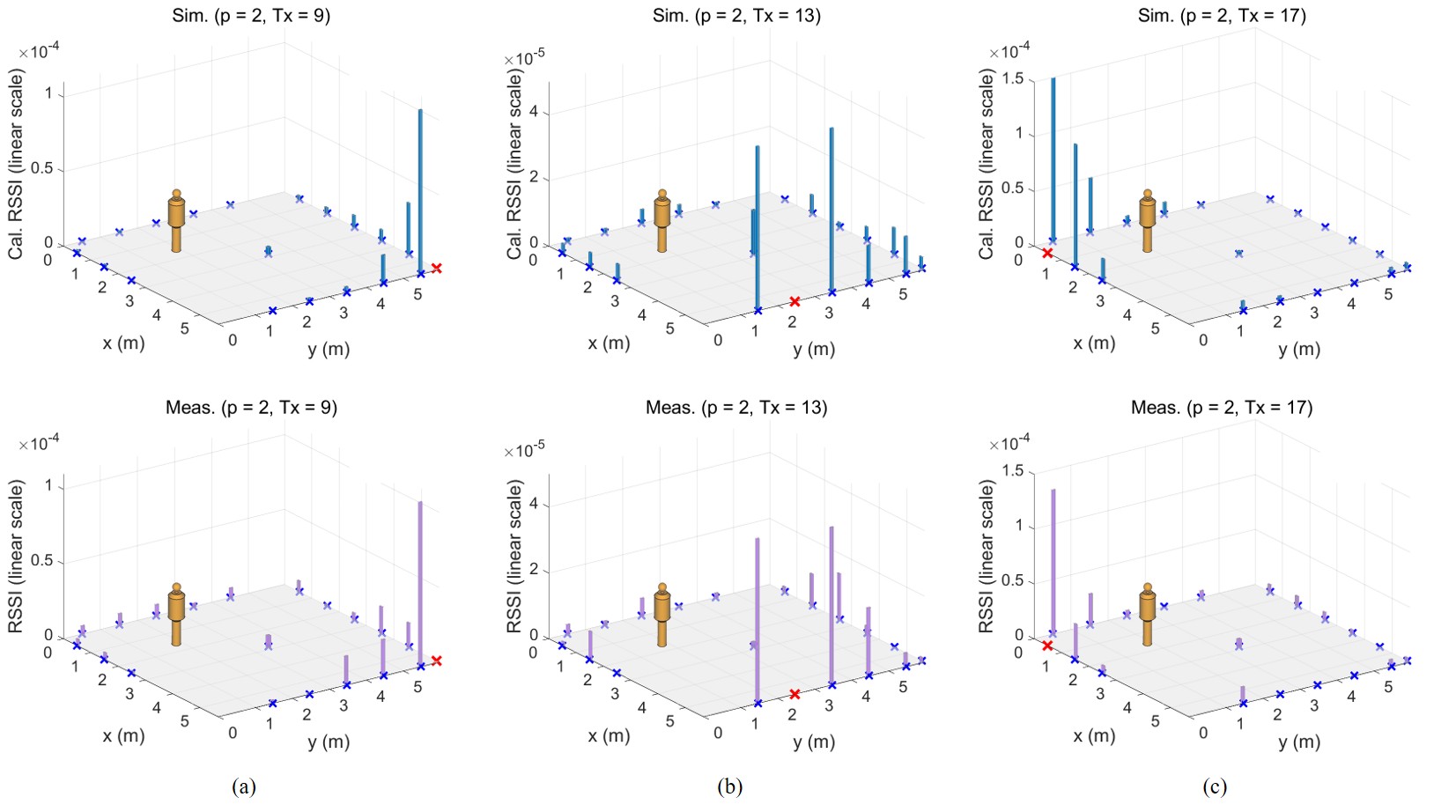}}
\caption{Comparison of measured and calibrated computational RSSI in linear scale across receivers for different transmitters with the simulated human body placed in $p=2$: (a) Tx = 9, (b) Tx = 13, and (c) Tx = 17. The transmitter is indicated by a red cross, the receivers by blue crosses, and the human model and its location are annotated in the figure.}
\label{fig.vali_linear}
\end{figure*}

The scattered and total $E_z$ fields on the $\rho,z$ plane and $x,y$ plane are illustrated in Fig.~\ref{fig.Ez_rz_xy}, in this case computed by the equivalent sources, since the observation region extends beyond the FEM domain. Fig.~\ref{fig.Ez_rz_xy}(a) and (b) show the scattered and total fields on the $\rho,z$ plane ($\phi = \phi_s$). As a further check on the accuracy of the results obtained, we temporarily extended the FEM domain to a large region around the body, and the field obtained was observed to match extremely well with that in Fig.~\ref{fig.Ez_rz_xy}(a) and (b), which have been obtained by the equivalent sources.  Note that the dark regions in Fig.~\ref{fig.Ez_rz_xy} correspond to the human body position, and the field is therefore not computed in that region (the equivalent sources yield zero field in the interior of surface $S_\text{eq}$). The scattered field in Fig.~\ref{fig.Ez_rz_xy}(a) reveals distinct scattering responses of different parts of the body, underscoring the advantage of a structurally informed human model that is capable of taking into account with good accuracy the field pattern at different heights from the ground. From Fig.~\ref{fig.Ez_rz_xy}(b), there is a clear effect of the ground reflection, in the form of an interference pattern in the vertical direction.
 The scattered total $E_z$ fields on the $x,y$ plane at $z=1$ m, corresponding to the plane of the Tx and Rxs, are illustrated in Fig.~\ref{fig.Ez_rz_xy}(c) and (d). The field pattern around the human body indicates diffraction and local interference, demonstrating that in the shadow region, the field is not null but has non-negligible fluctuations in amplitude. 
 As shown in Fig.~\ref{fig.Ez_rz_xy}(d), there is a clear interference pattern due to reflection from the ground. The results computed by 2.5-D FEM provide a detailed characterization of how the human body affects each Tx-Rx link, which is critical for modeling in human localization scenarios.


\subsection{Critical comparison with measurements}

Experimental measurements are now critically compared with the corresponding numerical results obtained in this work. As a premise, it should be mentioned that several uncertainties affect the experimental data, such as multiple reflections from the lateral walls, floor and ceiling of the room (of unknown material), scattering from objects that are present in the room itself, imperfect calibration~\cite{chen2010RSSI} of the RSSI measurements from the sensors and physical differences from the sensors themselves, small errors in the position of the transmitters, the receivers and the human body. Under these conditions, an accurate comparison between measurements and simulations is not really meaningful, and a different metric must be used to evaluate the usefulness of the model when inserted in real-case scenarios. 

The distribution of the RF sensing network nodes is configured as illustrated in Fig.~\ref{fig.sys2} and previously described in ~\cite{Federica_SysSetup}. The network uses IoT radio devices operating at 2.43 GHz (channel 16) and following the IEEE 802.15.4 physical layer standard, which is commonly adopted in industrial applications~\cite{IEEE802}. The network employs a time-slotted access in which each RF node transmits a physical protocol data unit (beacon) during its assigned time slot, with a slot duration of 0.5 ms and a guard time of 0.15 ms. During each slot, nodes measure the RSSI of all incoming beacons. The collected samples are then sent to a central access point (AP) that serves as a data aggregation and processing hub. Snapshots, consisting of RSSI measurements for all links in the network, are collected every 60 ms. The network is implemented using low-power NXP JN5189 SoC transceivers \cite{NXP_JN5189} and remains fully compatible with the IEEE 802.15.4 standard and common communication protocols such as  RFC 4944 (6LoWPAN) and TSCH (6TiSCH)~\cite{MultiModal}. Each node is equipped with a low-gain vertical monopole antenna. The test scenario is shown in Fig.~\ref{fig.sys2}(b).

During the measurements, nodes $N=1, \dots,20$ are sequentially selected as the transmitting node, while all remaining nodes operate as receivers. For each transmitting configuration, network data are collected with the human target positioned at five predefined locations, including natural body motions such as breathing, slight movements, and body rotations. RSSI measurements $P_{u,v}$, expressed in dBm, are collected by the network nodes as
\begin{equation}\label{eq:P_uv}
    P_{u,v} =
\begin{cases} 
P_{u,v}(\emptyset) + w_R, & p = 0\text{ (free-space)} \\
P_{u,v}(\emptyset) - A_{u,v} + w_T, & p =1, \cdots,5
\end{cases}
\end{equation}
where $u$ and $v$ denote the transmitting and receiving nodes, respectively, $P_{u,v}(\emptyset)$ is the RSSI in the free-space i.e. with no people inside the monitored area $(p=0)$, while $A_{u,v}$ is the extra attenuation (expressed in dB) due to the presence of the human body in the location $p$. $P_{u,v}(\emptyset)$ is assumed to be known or measured in the reference scenario ~\cite{Federica_SysSetup}. The noise terms $w_R$ and $w_T$ model the log-normal multipath fading and the other disturbances \cite{Fieramosca_awpl,Vittorio202201}.

 Since the practical radio devices have uncertain parameters such as transmit power, antenna gains, and system calibration, a transmitter-dependent scalar calibration factor is applied to the simulated RSSI to enable meaningful comparison. Specifically, the calibrated simulated RSSI $P_{u,v}^{\text{cal}}$ is computed as
\begin{equation}
    P_{u,v}^{\text{cal}}=\alpha_u P_{u,v}^{\text{sim}},
\end{equation}
where $\alpha_u$ is determined by matching the maximum simulated RSSI to the corresponding measured value for each transmitter. The experimental measurements correspond to the human subject located at $p=2$, for which a total of 251 snapshots were collected. Each snapshot consists of six consecutive samples, and the linear average of all samples is used for comparison.

The comparison between the calibrated simulated RSSI (blue lines) and the measured RSSI (violet lines) across all receiving nodes for three representative transmitting nodes is shown in Fig.~\ref{fig.vali_linear}. In this figure, we focus on three representative transmitting nodes, Tx = 9, 13, and 17, which cover a range of distances and relative positions with respect to the target human. Overall, the spatial distribution patterns for the simulated and measured RSSI are in good agreement. 

For the case of Tx = 9 shown on the left of Fig.~\ref{fig.vali_linear}(a), the transmitter is located at the farthest distance from the human target. RSSI peaks are observed at the receiving nodes closest to the transmitter, and both the simulation and measurement results capture the gradual decay of RSSI with increasing distance from the transmitter. The ripple distribution in the measured RSSI at nodes 15-20 reflects diffraction effects and local interference behind the human body. These effects are less pronounced in the simulated results, mainly due to multipath propagation and environmental reflections. 

\begin{figure*}[!t]
\centerline{\includegraphics[width=6.8in]{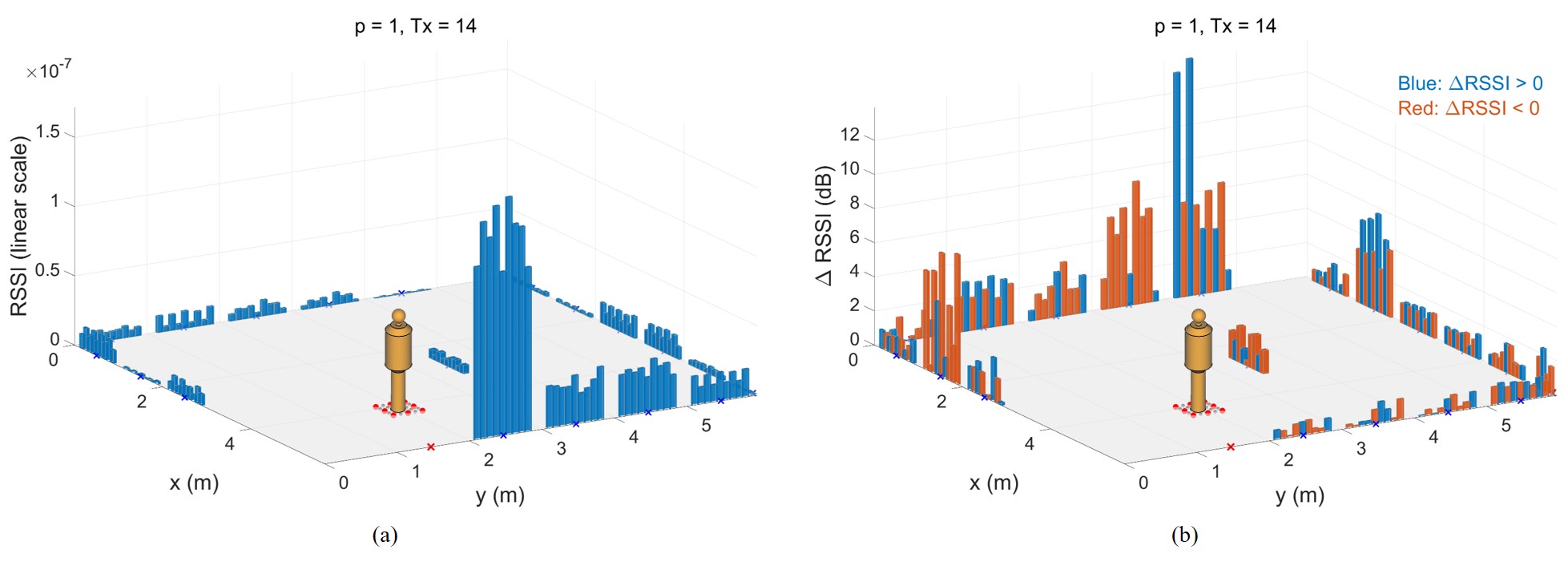}}
\caption{Effect of human micro-movements on the RSSI and $\Delta$RSSI for $p=1$ with Tx = 14. The transmitter is indicated by a red cross, the receivers by blue crosses, and the simulated human body and its location are shown in the figure as well. 5×5 sampled positions uniformly distributed around the human body within a 0.4 m × 0.4 m area are emphasized by gray circles, among which 3×3 representative positions, consisting of the four corner points, the four edge midpoints, and the center point of the grid, are selected for plotting and highlighted in red. The nine bars are ordered from left to right, which maps to points arranged with respect to the coordinate axes, ordered row-wise from left to right and from top to bottom. (a) Linear-domain RSSI, and (b) $\Delta$RSS in dB. Positive values are shown in blue, negative values in red for clarity.}
\label{fig.fem_micromove}
\end{figure*}

For the case of Tx = 13, shown in the center of Fig.~\ref{fig.vali_linear}(b), the transmitting device is at a moderate distance relative to the human body, and the RSSI distribution exhibits more spatial variations. Both the simulated and measured RSSI for devices 4-9 show noticeable fluctuations, indicating that both capture the human-body scattering as well as interference patterns resulting from the presence of the ground.

For the case of Tx = 19 in the right part of Fig.~\ref{fig.vali_linear}(c), the transmitter is very close to the human subject. The RSSI is dominated by a few nearby receivers, and the overall trend of the simulation results agrees well with the measurements. Compared to the measurements, the simulation slightly underestimates the RSSI, which may be attributed to additional scattering paths, noise, and reflections from walls and the ceiling present in the real environment.

In all cases, the calibrated simulated RSSI successfully captures the general spatial trends and relative amplitudes among the receiving nodes. These results are, however, only partially useful, since the uncertainties previously mentioned in the experimental setup may lead to severe differences in the received powers even in the presence of an accurate modelization. A more interesting metric is shown in the following section, where we analyze micro-movements of the human body.

\subsection{Analysis of Human Micro-Movements}

\begin{figure*}[!t]
\centerline{\includegraphics[width=7.1in]{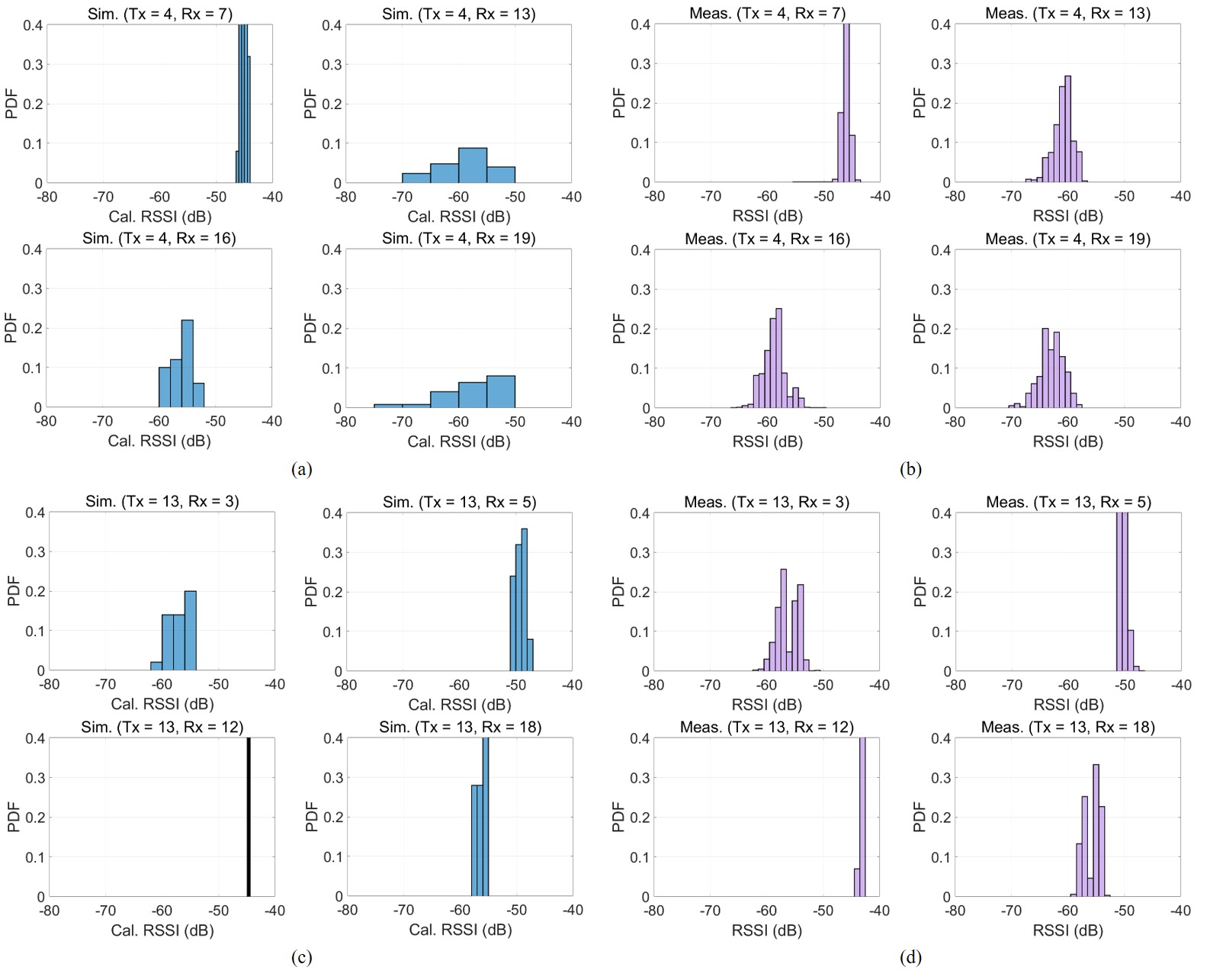}}
\caption{Comparison of single-link RSSI PDFs for $p=3$. For Tx = 4 with Rxs = 7, 13, 16, and 19: (a) calibrated calculated results, and (b) measured results; for Tx = 13 with Rxs = 3, 5, 12, and 18: (c) calibrated calculated results, and (d) measured results.}
\label{fig.vali_pdf}
\end{figure*}

In practical DFL and RF sensing applications, even slight movements of the target can significantly affect localization accuracy and system performance. Small displacements may alter scattering patterns, impacting the fidelity of received signal measurements and, consequently, the performance of sensing algorithms. Moreover, during the training of RF-based models, micro-movements are typically difficult to calibrate or label precisely, which can introduce uncertainties and reduce the effectiveness of the generated training data. Therefore, systematically studying and calibrating the effects of slight human displacements in simulation can be a good way to address this issue.

In this work, these small displacements are modeled by uniformly sampling 25 positions within a 0.4~m $\times$ 0.4~m area around the initial positions of the human target. Using the proposed 2.5-D FEM, the RSSI is calculated at each position to assess how small/involuntary changes in body position (i.e., micro-movements) influence the distribution of received signals. The concept of $\Delta$RSSI is introduced to quantitatively analyze these effects: this is defined as $\Delta\text{RSSI}_{u,v}=P_{u,v}(\emptyset) - P_{u,v}$, where $P_{u,v}(\emptyset)$ corresponds to the RSSI in the absence of a human, and $P_{u,v}$ corresponds to the RSSI with a human present, which are both definded in \eqref{eq:P_uv}. This differential measure effectively mitigates the influence of the static environment, isolating the impact of the human target on the sensing network. 

Transmitter 14 and $p=1$ are selected as an example to illustrate the impact of human micro-movements on the RSSI and $\Delta$RSSI, as shown in Fig.~\ref{fig.fem_micromove}. To facilitate visualization, only 3×3 points out of the 25 sampled positions are plotted. In Fig.~\ref{fig.fem_micromove}(a), micro-movements of the human target do not introduce significant changes in the relative RSSI distribution across the receiving nodes. The RSSI peak consistently appears at the receiver closest to the transmitting node, while the overall spatial distribution among the remaining receivers remains largely consistent with the trends observed in previous analyses. This indicates that linear-domain RSSI exhibits limited sensitivity to micro-movements of the target. 

In contrast, the $\Delta$RSSI shown in Fig.~\ref{fig.fem_micromove}(b) demonstrates a markedly stronger response to micro-movements. Depending on the direction and position of the movement, both the magnitude and the sign of $\Delta$RSSI vary significantly. This effect is particularly pronounced at receivers located behind the human body with respect to the transmitter. Since these receivers are farther from the transmitting node, their received signals are more sensitive to perturbations induced by human scattering and shadowing. The observed variations in $\Delta$RSSI values and polarity across different displacement positions essentially reflect the fact that the human body interacts with distinct local field distributions as its position changes.

This implies that when $\Delta$RSSI is employed as a training feature in RF sensing networks, uncalibrated micro-movements of human targets may introduce uncertainties into experimental datasets, potentially degrading training performance and localization accuracy. In this context, the proposed method enables the generation of calibrated RF databases that explicitly account for small-scale human displacements, thereby facilitating more accurate and physically informed learning-based DFL and RF sensing models.

To further validate the reliability of the proposed method on micro-movement analysis, Fig.~\ref{fig.vali_pdf} compares the probability density functions (PDFs) of RSSI (in dB) for individual links between the measured and calculated results. We select $p=3$ as an example. The results include Tx = 4, which is close to the human, and Tx = 13, which is relatively farther, with four different receiving nodes selected for different distances from each transmitter to illustrate the link-specific RSSI distributions. The calculated data are obtained from 25 sampled positions representing micro-movements around the initial human location, while all of the measured results for $p=3$ are used, since they include unavoidable slight movements and rotations of the human during the experiment. As in the previous processing, the calculated results are calibrated to match the range of the measured RSSI.

From the results shown in Fig.~\ref{fig.vali_pdf}, it can be observed that the calculated and measured RSSI PDFs exhibit similar trends across different links, including the central tendency, main fluctuation ranges, and spread of the distributions. Moreover, the simulation effectively reflects the relative differences among links. The variations in RSSI PDF observed across different receiving antennas under an identical transmitting antenna are successfully captured by the simulation. This indicates that the proposed method not only predicts the average RSSI levels at individual receivers but also captures the main statistical fluctuations induced by slight human displacements across different links.

Although perfect agreement between simulation and measurement is not to be expected due to the complexity of the experimental environment, to the uncertainties of micro-movements of human targets, and to the differences in sample size, the overall trends and inter-link variations are well preserved. These results further demonstrate the reliability of the proposed 2.5-D FEM framework in simulating the impact of micro-movements on RSSI statistics and its applicability for constructing labeled RF databases to support DFL and RF sensing applications.

\section{Conclusion}
\label{chap:conclusion}

In this paper, a fast 2.5-D FEM is presented for computing the scattering fields of a BoR human body model under the excitation of a z-directed dipole. The numerical accuracy of the proposed method is validated through comparisons with full-wave 3-D simulations using Feko$^\text{TM}$. Additionally, in indoor DFL scenarios, simulated RSSI values show reasonable agreement with experimental measurements, demonstrating the applicability of the method for predicting spatial RSSI distributions. The impact of human micro-movements on RSSI statistical characteristics is also analyzed. The results indicate that the proposed method can partially reproduce the PDFs of RSSI variations, capturing both the mean trends and statistical fluctuations observed experimentally. Overall, the proposed 2.5-D FEM significantly reduces computational complexity compared to conventional 3-D solvers, while preserving the essential geometric features of the human body and providing accurate and statistically consistent results. This makes it a practical tool for generating large-scale, calibrated RF datasets, supporting the rapid development and evaluation of physics-prior models for indoor DFL and RF sensing applications.

Future work will focus on extending the proposed 2.5-D FEM in two directions. First, the method will be further developed for directional antennas, aiming to enhance the realism of computed scenarios. Second, we will investigate the extension of the method to multi-target environments, establishing interaction models among multiple human bodies to strengthen the capability for more complex and realistic RF simulations of the tool. These developments are expected to enable more comprehensive studies of electromagnetic interactions in complex indoor environments.

\appendices
\section{Matrix Elements for the 2.5-D FEM}
\label{app:matrix_element}

In this appendix, we show the matrix elements in \eqref{eq:Amat} and the following equations. Letting
\begin{equation}
    (\nabla\times{\bm\tau})_t = {\tau_z\irho -\tau_\rho\iz},
\end{equation}
\begin{equation}
    (\nabla\times{\bm\tau})_\phi = \frac{\partial\tau_{\rho}}{\partial z}-\frac{\partial \tau_{z}}{\partial \rho},
\end{equation}
\begin{equation}
    (\nabla\times\iphi \varphi)_t = \frac{\partial \varphi}{\partial \rho}\iz -\frac{\partial \varphi}{\partial z}\irho,
\end{equation}
we have, indicating with $\mathbb{M}(p,q)$ element $p,q$ of matrix $\mathbb{M}$
\begin{equation}
    \mathbb{A}_{tt}(p,q) = \int_S (\nabla\times{\bm\tau}_p)_\phi(\nabla\times{\bm\tau}_q)_\phi \;\rho d\rho dz,
\end{equation}
\begin{equation}
    \mathbb{A}'_{tt}(p,q) = \int_S (\nabla\times{\bm\tau}_p)_t\cdot(\nabla\times{\bm\tau}_q)_t \frac{1}{\rho} d\rho dz,
\end{equation}
\begin{equation}
    \mathbb{A}_{t\phi}(p,q) = \int_S (\nabla\times{\bm\tau}_p)_t\cdot(\nabla\times\iphi\varphi_q)_t \frac{1}{\rho} d\rho dz,
\end{equation}
\begin{equation}
    \mathbb{A}_{t\phi}(p,q) = \int_S (\nabla\times\iphi\varphi_p)_t\cdot(\nabla\times{\bm\tau}_q)_t \frac{1}{\rho} d\rho dz,
\end{equation}
\begin{equation}
    \mathbb{A}_{\phi\phi}(p,q) = \int_S (\nabla\times\iphi\varphi_p)_t\cdot(\nabla\times\iphi\varphi_q)_t \frac{1}{\rho} d\rho dz,
\end{equation}
\begin{equation}
    \mathbb{B}_{tt}(p,q) = \int_S \epsilon_r {\bm\tau}_p\cdot {\bm\tau}_q \;\rho d\rho dz,
\end{equation}
\begin{equation}
    \mathbb{B}_{\phi\phi}(p,q) = \int_S \epsilon_r \frac{{\varphi}_p {\varphi}_q}{\rho} \; d\rho dz.
\end{equation}
As for matrix $\mathbb{C}$, imposing an impedance-type boundary condition
\begin{equation}
    \normal\times{\bf E} = \zeta\normal\times{\bf H}\times\normal,
\end{equation}
one finds
\begin{equation}
    \mathbb{C}_{tt}(p,q) = \zeta\int_{\partial S_\text{gnd}} {\bm\tau}_p\cdot (\normal\times{\bm\tau}_q\times\normal) \; \rho d\rho dz ,
\end{equation}
\begin{equation}
    \mathbb{C}_{\phi\phi}(p,q) = \zeta\int_{\partial S_\text{gnd}} \frac{{\varphi}_p {\varphi}_q}{\rho} \; d\rho dz,
\end{equation}
where ${\partial S_\text{gnd}}$ is the line corresponding to the ground surface in $\rho,z$ plane.

\bibliographystyle{ieeetr}
\bibliography{ref}

\end{document}